\documentstyle[sprocl,epsfig]{article}

\def\be{\begin{equation}}
\def\ee{\end{equation}}
\def\bea{\begin{eqnarray}}
\def\eea{\end{eqnarray}}

\begin{document}

\title{Supernova Studies at ORLaND}

\author{A. MEZZACAPPA}

\address{Physics Division, Oak Ridge National Laboratory\\ Bldg. 6010, 
MS 6354, P.O. Box 2008\\ Oak Ridge, TN 37831-6354\\ E-Mail:
mezzacappaa@ornl.gov} 


\maketitle\abstracts{
A new facility to measure neutrino mass differences and mixing angles 
and neutrino--nucleus cross sections, such as the proposed ORLaND facility 
at Oak Ridge, would contribute to the experimental determination of vacuum 
mixing parameters and would provide an experimental foundation for the many 
neutrino--nucleus weak interaction rates needed in supernova models. This 
would enable more realistic supernova models and a far greater ability 
to cull fundamental physics from these models by comparing them with 
detailed observations. 
Charged- and neutral-current neutrino interactions on nuclei in the stellar 
core play a central role in supernova dynamics, nucleosynthesis, and neutrino 
detection. Measurements of these reactions on select, judiciously chosen targets 
would provide an invaluable test of the complex theoretical models used to compute 
the neutrino--nucleus cross sections.
}

\section{Introduction}
Core collapse supernovae are among the most energetic explosions in the 
Universe, releasing $10^{53}$ erg of energy in the form of neutrinos of 
all flavors at the staggering rate of $10^{57}$ neutrinos per second and 
$10^{45}$ Watts, disrupting almost entirely stars more massive than 8--10 
M$_{\odot}$ and producing and disseminating into the interstellar medium 
many of the elements in the periodic table. 
They are a key link in our chain of origins from the Big Bang
to the formation of life on Earth; a nexus of nuclear physics, particle 
physics, fluid dynamics, radiation transport, and general relativity; 
and serve as laboratories for physics beyond the Standard Model and 
for matter at extremes of density, temperature, and neutronization 
that cannot be produced in terrestrial laboratories. 

Current supernova theory centers around the idea that the supernova shock 
wave---formed when the iron core of a massive 
star collapses gravitationally and rebounds as the core
matter exceeds nuclear densities and becomes incompressible---stalls 
in the iron core as a result of enervating losses to nuclear dissociation 
and neutrinos. The failure of this ``prompt'' supernova mechanism 
sets the stage for a ``delayed'' mechanism, whereby the shock is reenergized 
by the intense neutrino flux emerging from the neutrinospheres carrying 
off the binding energy of the 
proto-neutron star\cite{w85,bw85}. The heating is mediated primarily by the 
absorption of electron neutrinos and antineutrinos on the dissociation-liberated 
nucleons behind the shock. 
This process depends critically on 
the neutrino luminosities, spectra, and angular distributions, 
i.e., on the multigroup (multi-neutrino energy) neutrino transport 
between the proto-neutron star and the shock.
This past decade has also seen the emergence 
of multidimensional supernova models, which have investigated the 
role convection, rotation, and magnetic fields may play in the explosion
\cite{hbc92,mwm93,hbhfc94,bhf95,jm96,mcbbgsu98a,mcbbgsu98b,s98,fh99,khowc99}.

Realistic supernova models will require extremely accurate neutrino radiation
hydrodynamics, but unless advancements in neutrino transport 
are matched by equally important
advancements in nuclear and weak-interaction physics, the efficacy
of the former must be called into question. In particular, we must move
forward to the use of ensembles of nuclei in the stellar core rather 
than a single representative nucleus when computing electron and 
electron neutrino capture during the critical core collapse phase. 
The use of a single representative nucleus has been the standard 
until now in virtually all supernova models. Moreover, in computing 
the electron and electron neutrino capture rates, and in general all 
of the neutrino--nucleus cross sections, detailed shell model computations
must replace the parameterized approximations that have been used
in the past. For example, recent work on electron capture up to mass 
65 has shown that these parameterized rates can be orders of magnitude 
in error\cite{lmp00}. The electron capture rate is dominated by Gamow--Teller 
resonance transitions, with the Gamow--Teller strength distributed over 
many states. Parameterized treatments place the resonance at a single
energy, and this energy is often grossly over- or underestimated
relative to the Gamow--Teller centroid computed in realistic shell 
model computations, leading in turn to rates that are often grossly 
too small or too large, respectively. (If the centroid is 
underestimated, more electrons can participate in capture.
The reverse is true if the centroid is too high.)

Whereas generations of nuclear structure models will afford 
ever greater realism in the calculation of stellar core properties
and the interactions of core nuclei with the neutrinos flowing through 
the core, nuclear
experiments must be designed and carried out that will serve as
guide posts for the theoretical predictions that must be 
made for the countless rates that enter into any realistic supernova 
or supernova nucleosynthesis model. In particular, we must have 
neutrino--nucleus cross section measurements that will help gauge 
neutrino capture and scattering predictions during stellar core 
collapse and during the p-, r- and $\nu$-processes after core bounce.

The plan of this paper is as follows. First we give an overview of 
the state of the art in one- and two-dimensional supernova models,
confining our discussion to issues of neutrino transport and 
convection. For discussions of the role of general relativity, 
rotation, and magnetic fields in supernova models, the reader 
may begin with the papers by Bruenn {\it et al.}\cite{bdm00}, 
Liebend\"{o}rfer {\it et al.}\cite{l00},\cite{lmtmhb00}, Fryer and Heger\cite{fh99},
Khokhlov {\it et al.}\cite{khowc99}, and MacFadyen and Woosley\cite{mw99}. 
Our focus will then turn to the nuclear and neutrino science that is 
input to these models and important for supernova neutrino detection,
with an eye toward measurements that could be made at a stopped-pion
facility such as ORLaND.

\section{One-Dimensional Supernova Models}
Although three decades of supernova modeling have established a theoretical
framework, fundamental questions about the explosion mechanism remain. Is the 
neutrino heating sufficient, or are multidimensional effects such as convection 
and rotation necessary? Can the basic supernova observable, explosion, be 
reproduced by detailed spherically symmetric models, or are multidimensional 
models required? In all of their phenomenology, core collapse supernovae 
are not spherically symmetric. For example, neutron star kicks\cite{fbb98} 
and the polarization of supernova emitted light\cite{w99} cannot arise in 
spherical symmetry. Nonetheless, ascertaining the explosion mechanism
and understanding every explosion observable are two different goals. 
To achieve both, simulations in one, two, and three dimensions must be 
coordinated.

The neutrino energy deposition behind the shock depends sensitively 
on the neutrino luminosities, spectra, and angular distributions in 
the postshock region. Ten percent variations in any of these quantities 
can make the difference between explosion and failure in supernova 
models\cite{jm96,bg93}. Thus, exact multigroup Boltzmann neutrino 
transport must be considered in supernova models.
Past spherically symmetric simulations have implemented increasingly 
sophisticated approximations to Boltzmann transport: simple leakage 
schemes\cite{vl81}, two-fluid models\cite{cvb86}, and multigroup 
flux-limited diffusion\cite{ar77,br93,wm93}. A generic feature of 
this last, most sophisticated approximation is that it underestimates 
the isotropy of the neutrino angular distributions in the heating region 
and, thus, the heating rate\cite{ja92,mmbg98}. Failure to produce 
explosions in the past may have resulted from the use of transport
approximations.

To address this question, we model the core collapse, bounce, and 
postbounce evolution of a 13 M$_{\odot}$ star, beginning with the
precollapse model of Nomoto and Hashimoto\cite{nh88}, with a new 
neutrino radiation hydrodynamics code for both Newtonian and general 
relativistic spherically symmetric flows: AGILE--BOLTZTRAN.
BOLTZTRAN is a three-flavor Boltzmann neutrino transport 
solver\cite{mb93b,mm99}, now extended to fully general
relativistic flows\cite{l00}. In the simulation we 
include here\cite{mlmhtb00}, it was employed in the 
$O(v/c)$ limit. AGILE is a conservative general relativistic 
hydrodynamics code\cite{l00,lt98}. Its adaptivity enables us 
to resolve and seamlessly follow the shock through the iron 
core into the outer stellar layers.

\begin{figure}
\begin{center}
\epsfysize=2.5in 
\epsfbox{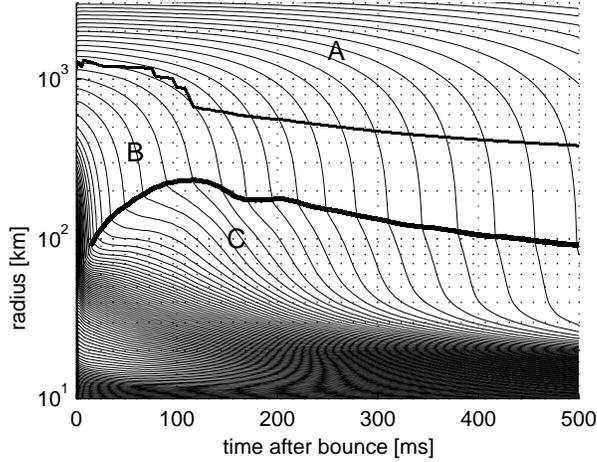} 
\caption[h]{
We trace the shock, nuclear burning, and dissociation 
fronts (the shock and dissociation fronts are coincident), 
which carve out three regions in the $(r,t)$ plane. 
A: Silicon. 
B: Iron produced by infall compression and heating. 
C: Free nucleons and alpha particles. 
}
\end{center}
\end{figure}

Figure 1, taken from the simulation of Mezzacappa {\it et al.}\cite{mlmhtb00} 
shows the radius-versus-time trajectories of equal mass (0.01M$_{\odot}$) 
shells in the stellar iron core and silicon layer in our Newtonian simulation. 
Core bounce and the formation and propagation of the initial bounce 
shock are evident. This shock becomes an accretion shock, decelerating 
the core material passing through it. At $\sim$ 100 ms after bounce, 
the accretion shock stalls at a radius $\sim$ 250 km and 
begins to recede, continuing to do so over the next several
hundred milliseconds. No explosion has developed in this model during 
the first $\sim$ 500 ms.

Thus, we are beginning to answer some fundamental questions in supernova 
theory. We have shown results from the first $\sim$ 500 ms of our Newtonian 
core collapse supernova simulation with Boltzmann neutrino transport, initiated 
from a 13 M$_{\odot}$ progenitor. In light of our implementation of Boltzmann 
transport, if we do not obtain explosions in this model or its general 
relativistic counterpart when they are completed, or in subsequent models 
initiated from different progenitors, it would suggest that either changes in our 
initial conditions (precollapse models) and/or input physics or the inclusion 
of multidimensional effects such as convection, rotation, and magnetic fields 
are required ingredients in the recipe for explosion. With the implementation 
of Boltzmann transport, this conclusion can be made unambiguously. In the past, 
it was not clear whether failure or success in supernova models was the result 
of inadequate transport approximations or the lack of inclusion of important 
physics.

With regard to improved input physics, the use of ensembles of 
nuclei in the stellar core rather than a single representative 
nucleus, computing the neutrino--nucleus cross sections with 
detailed shell model computations\cite{lmp00}, the inclusion 
of nucleon correlations in the high-density neutrino 
opacities\cite{bs98,rplp99}, and improvements in 
precollapse models\cite{ba98,unn99} all have the 
potential to quantitatively, if not qualitatively, 
change the details of our simulations. Thus, it is 
important to note that our conclusions are drawn 
in the context of the best available input physics.

\section{Two-Dimensional Supernova Models: Convection}

Supernova convection falls into two categories: (1) convection 
near or below the neutrinospheres, which we refer to as proto-neutron
star convection and (2) convection between the gain radius and the
shock, which we refer to as neutrino-driven convection. Proto-neutron
star convection may aid the explosion mechanism by boosting the 
neutrinosphere luminosities, transporting by convection hot, 
lepton-rich rich matter to the neutrinospheres. Neutrino-driven
convection may aid the explosion mechanism by boosting the shock
radius and the
neutrino heating efficiency, thereby facilitating shock revival.

\subsection{Proto-Neutron Star Convection}
This mode of convection may develop owing to instabilities caused by 
lepton and entropy gradients established by the deleptonization of the 
proto-neutron star via electron neutrino escape near the electron neutrinosphere 
and by the weakening supernova shock. (As the shock weakens, it causes
a smaller entropy jump in the material flowing through it.) Proto-
neutron star convection is arguably the most difficult to investigate 
numerically because the neutrinos and the matter are coupled, and, 
consequently, multidimensional simulations must include both multidimensional 
hydrodynamics and multidimensional, multigroup neutrino transport.

In certain regions of the stellar core, neutrino transport can equilibrate 
a convecting fluid element with its surroundings in both entropy and lepton 
number on time scales shorter than convection time scales, rendering 
the fluid element nonbouyant. This will occur in intermediate regimes 
in which neutrino transport is efficient but in which the neutrinos are still 
strongly enough coupled to the matter. Figures 2 and 3, from Mezzacappa 
{\it et al.}\cite{mcbbgsu98a}, demonstrate that this equilibration can in fact 
occur. Figure 2 shows the onset and development of proto-neutron star 
convection in a 25 M$_{\odot}$ model shortly after bounce in a simulation 
that did not include neutrino transport, i.e., that was a hydrodynamics-only 
run. Figure 3 on the other hand shows the lack of any significant onset and 
development of convection when neutrino transport was included in what was 
otherwise an identical model. Transport's damping effects are obvious. (The 
same result occurred in our 15 M$_{\odot}$ model.) 

On the other hand, in the model of Keil {\it et al.}\cite{kjm96}, vigorous proto-neutron 
star convection developed, which then extended deep into the core as a 
deleptonization wave moved inward, owing to neutrinos diffusing outward. 
In this model, convection occurs very deep in the core where neutrino 
opacities are high and transport becomes inefficient in equilibrating 
a fluid element with its surroundings.

It is important to note in this context that Mezzacappa {\it et al.}
and Keil {\it et al.} used complementary transport approximations. In
the former case, spherically symmetric transport was used, which
maximizes lateral neutrino transport and overestimates the 
neutrino--matter equilibration rate; in the latter case, ray-by-ray
transport was used, which minimizes (zeroes) lateral transport and
underestimates the neutrino--matter equilibration rate.

These different outcomes clearly demonstrate that to determine whether 
or not 
proto-neutron star convection exists and, if it exists, is vigorous will 
require simulations coupling three-dimensional, multigroup neutrino transport 
and three-dimensional hydrodynamics. Moreover, realistic high-density neutrino 
opacities will be needed. 

\begin{figure}
\epsfxsize=4.7in 
\hspace{4.0cm}\epsfbox{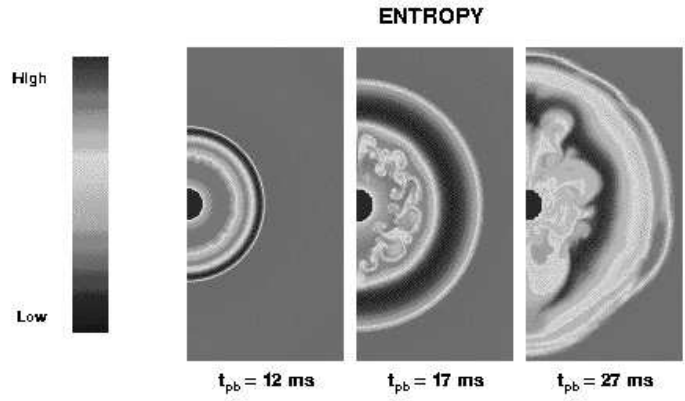} 
\caption[h]{Two-dimensional entropy plots showing the evolution of
proto-neutron star convection in our hydrodynamics-only 25 ${\rm M}_{\odot}$ 
model at 12, 17, and 27 ms after bounce.}
\end{figure}

\begin{figure}
\epsfxsize=4.7in 
\hspace{4.0cm}\epsfbox{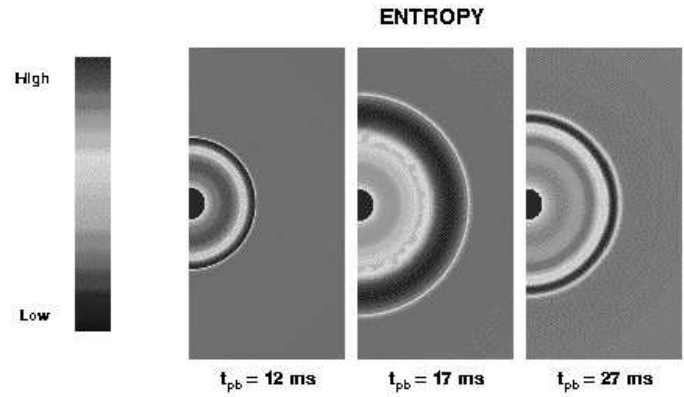} 
\caption[h]{Two-dimensional entropy plots showing the evolution of
proto-neutron star convection in our hydrodynamics-plus-neutrino-transport 
25 ${\rm M}_{\odot}$ model at 12, 17, and 27 ms after bounce.}
\end{figure}
 
\subsection{Neutrino-Driven Convection}
This mode of convection occurs directly between the gain
radius and the stalled shock as a result of the entropy 
gradient that forms as material infalls between the 
two while being continually heated. In Figure 5, a sequence of
two-dimensional plots of entropy are shown, illustrating
the development and evolution of neutrino-driven convection
in our 15 M$_{\odot}$ model\cite{mcbbgsu98b}. High-entropy, rising
plumes and lower-entropy, denser, finger-like downflows are seen.
The shock is distorted by this convective activity.

\begin{figure}
\epsfxsize=4.7in 
\hspace{4.0cm}\epsfbox{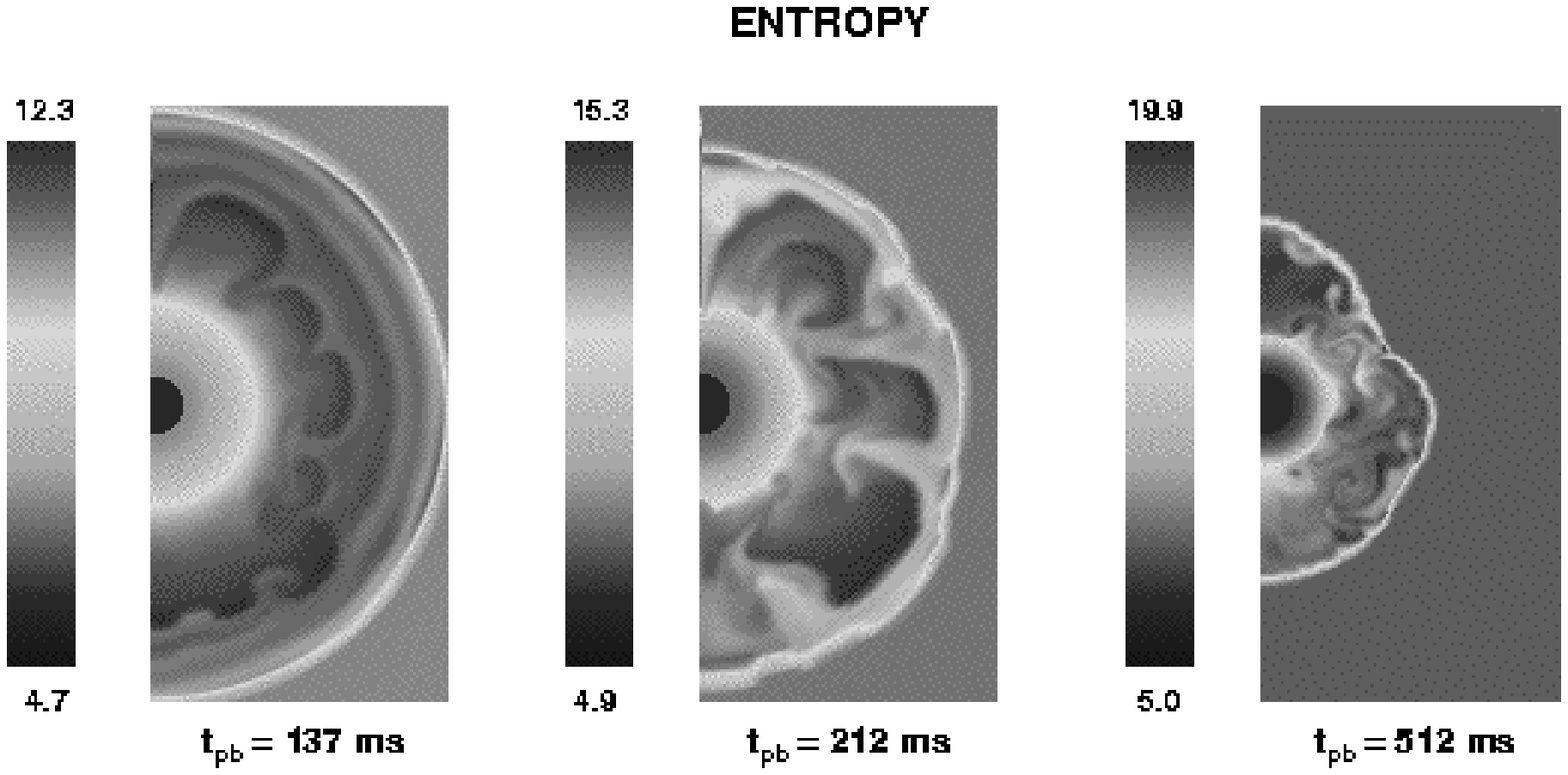} 
\caption[h]{Two-dimensional entropy plots showing the evolution of
neutrino-driven convection in our 15 ${\rm M}_{\odot}$ model
at 137, 212, and 512 ms after bounce.}
\end{figure}

In the Herant {\it et al.}\cite{hbhfc94} simulations, large-scale 
convection developed beneath the shock, leading to increased neutrino 
energy deposition, the accumulation of mass and energy in the gain 
region, and a thermodynamic engine that ensured explosion, although
Herant {\it et al.} stressed the need for more sophisticated multidimensional, 
multigroup transport in future models. [They used two-dimensional ``gray''
(neutrino-energy--integrated, as opposed to multigroup) flux-limited
diffusion in neutrino-thick regions and a neutrino lightbulb approximation 
in neutrino-thin regions. In a lightbulb approximation, the neutrino 
luminosities and rms energies are assumed constant with radius.]
In the Burrows {\it et al.} simulations\cite{bhf95}, neutrino-driven convection 
in some models significantly boosted the shock radius and led to explosions. 
However, they stressed that success or failure in producing explosions was 
ultimately determined by the values chosen for the neutrino spectral parameters 
in their gray ray-by-ray (one-dimensional) neutrino diffusion scheme. (In 
spherical symmetry (1D), all rays are the same. In a ray-by-ray 
scheme in axisymmetry (2D), not 
all rays are the same, although the transport along each ray is a 1D problem. 
In the latter case, lateral transport between rays is ignored.) 
Focusing on the neutrino luminosities, Janka and M\"{u}ller\cite{jm96}, using 
a central adjustable neutrino lightbulb, conducted a parameter survey and 
concluded that neutrino-driven convection aids explosion only in a narrow 
luminosity window ($\pm 10\%$), below which the luminosities are too low 
to power explosions and above which neutrino-driven convection is not necessary. 
In more recent simulations carried out by Swesty\cite{s98} using two-dimensional 
gray flux-limited diffusion in both neutrino-thick and neutrino-thin regions, it 
was demonstrated that the simulation outcome varied dramatically as the 
matter--neutrino ``decoupling point,'' which in turn sets the neutrino spectra 
in the heating region, was varied within reasonable limits. (The 
fundamental problem in gray transport schemes is that the neutrino 
spectra, which are needed for the heating rate, are not computed. 
The spectra are specified by choosing a neutrino ``temperature,''
normally chosen to be the matter temperature at decoupling. In a
multigroup scheme, the spectra are by definition computed.)  
In our two-dimensional models, the angle-averaged shock radii 
do not differ significantly from the shock trajectories in 
their one-dimensional counterparts, and no explosions are 
obtained, as seen in Figure 6. Neither the luminosities nor 
the neutrino spectra are free parameters. Our two-dimensional 
simulations implemented spherically symmetric (1D) multigroup 
flux-limited diffusion neutrino transport, compromising transport 
dimensionality to implement multigroup transport and a seamless 
transition between neutrino-thick and neutrino-thin regions.

In light of the neutrino transport approximations made, the fact that 
all of the simulations have either been one- or two-dimensional, and the mixed 
outcomes, next-generation simulations will have to reexplore neutrino-driven 
convection in the context of three-dimensional simulations that implement 
more realistic multigroup three-dimensional neutrino transport.

\section{\bf Neutrino--Nucleus Cross Sections}

Neutrino--nucleus cross sections of relevance to supernova
astrophysics fall into three categories: cross sections 
for (1) supernova dynamics, (2) supernova nucleosynthesis,
and (3) terrestrial supernova neutrino detection.

\subsection{\bf Supernova Dynamics}

Whether or not a supernova occurs is set at the time 
the shock forms and the entire post--stellar-core-bounce evolution 
is set in motion. Where the shock forms in the stellar core at bounce 
and how much energy it has initially are set by the ``deleptonization'' 
of the core during collapse. The deleptonization occurs as electrons
are captured on the free protons and iron-group nuclei in the core, 
producing electron neutrinos that initially escape. Deleptonization 
would be complete if electron capture continued without competition,
but at densities of order $10^{11-12}$ g/cm$^{3}$, the electron
neutrinos become ``trapped'' in the core, and the inverse 
reactions---charged-current electron neutrino capture on neutrons and 
iron-group nuclei---begin to compete with electron capture until 
the reactions are in weak equilibrium and the net deleptonization 
of the core ceases on a core collapse time scale. The equilibration 
of electron neutrinos with the stellar core occurs at densities 
between $10^{12-13}$ g/cm$^{3}$. Additionally, as the stellar 
core densities increase, the characteristic nuclei in the core 
increase in mass, owing to a competition between Coulomb contributions
to the nuclear free energy and nuclear surface tension. 
For densities of order $10^{13}$ g/cm$^{3}$, the nuclear mass is 
of order 140. Thus, cross sections for charged-current electron neutrino capture 
on iron-group nuclei through mass 100 are needed to accurately simulate 
core deleptonization and to accurately determine the postbounce initial 
conditions.

Table 1 summarizes the thermodynamic conditions in the core at 
the three densities discussed above and gives the representative 
nuclear mass and charge and mean electron neutrino energy.
The data were taken from a core collapse simulation carried
out by Mezzacappa and Bruenn\cite{mb93a,mb93c}. Electron neutrino 
capture would remain important until the neutrinos equilibrate 
with the matter, which in our simulation would occur when the 
representative nucleus in the core is between mass 88 and 138.

\begin{table}
\begin{center}
\begin{tabular}{||c|c|c|c|c|c||}
\hline
$\rho$ (g/cm$^{3}$) & Y$_{\rm e}$ & T (MeV) & A   & Z  & $<\epsilon_{\nu_{\rm e}}>$ (MeV) \\
\hline
\hline
$10^{11}$            & 0.4         & 1      & 70  & 30 & 12 \\
$10^{12}$            & 0.35        & 2      & 88  & 36 & 19 \\
$10^{13}$            & 0.3         & 4      & 138 & 52 & 43 \\
\hline
\end{tabular}
\end{center}
\end{table}

The size of the inner, unshocked core is proportional to $<Y_{\rm e}>^{2}$, 
where $<Y_{\rm e}>$ is the {\it mean} electron fraction in the inner core. 
Moreover, the shock loses $\sim 10^{51}$ erg of energy (an explosion energy) 
for every 0.1 M$_{\odot}$ it dissociates, which is $\sim$10--20\% of the total 
inner core mass. Thus, an $\sim$5--10\% change in the mean electron fraction 
would have a significant impact on the postbounce evolution. The 
mean electron fraction at bounce results from many capture reactions during 
infall (on both protons and nuclei; we focus on nuclei here), 
and it is clear that accurate electron and 
electron neutrino capture rates are needed and that theory must be checked 
against experiment even if only in a few strategic cases. 

One goal of the proposed ORLaND facility will be to measure the cross 
section for electron neutrino charged-current capture on $^{56}$Fe:

\begin{itemize}
\item $^{56}{\rm Fe}(\nu_{\rm e},e^{-})^{56}Co$
\end{itemize}
 
\noindent Pioneering measurements of this cross section have been 
performed by the KARMEN collaboration with an 
experimental uncertainty $\sim$ 50\%. Further measurements are 
required to achieve an accuracy $\sim$ 10\% to adequately test
theoretical models.
Moreover, the same proposed technique to measure this cross section
can be used to measure the electron neutrino capture cross section on any of 
the following nuclei, several of which are in the critical nuclear mass range 
mentioned above: $^{7}$Li, $^{9}$Be, $^{11}$B, $^{27}$Al, $^{40}$Ca, $^{51}$V, 
$^{52}$Cr, $^{55}$Mn, $^{59}$Co, $^{93}$Nb, $^{115}$In, $^{181}$Ta, and 
$^{209}$Bi.

\subsection{\bf Supernova Nucleosynthesis}

There are several ``processes'' that define supernova nucleosynthesis:
(1) Explosive nucleosynthesis, which occurs as a result of compressional
heating by the supernova shock wave as it passes through the stellar 
layers. (2) Neutrino nucleosynthesis or a ``neutrino process,'' which occurs due to nuclear 
transmutations in the stellar layers prior to shock passage. (3) A 
rapid neutron capture or ``r'' process, which occurs in the neutrino-driven 
wind that emanates from the proto-neutron star after the explosion is 
initiated. The neutrinos both drive the wind and interact with the
nuclei in it. Moreover, transmutations produced in (2) are postprocessed 
in (1). Thus, neutrino--nucleus interactions are central to all three 
nucleosynthesis processes, although indirectly to process (1).

\begin{center}
{\it Neutrino Nucleosynthesis}
\end{center}

Neutrino nucleosynthesis is driven by the spallation of
protons, neutrons, and alpha particles from nuclei in the 
stellar layers by the intense neutrino flux that is emanating 
from the central proto-neutron star powering the supernova\cite{whhh90}. 
Moreover, neutrino nucleosynthesis continues after the initial 
inelastic scattering reactions and the formation of their 
spallation products. The neutrons, protons, and alpha particles 
released continue the nucleosynthesis through further reactions 
with other abundant nuclei in the high-temperature supernova 
environment, generating new rare species. Neutrino nucleosynthesis 
occurs in two stages: (1) through the neutrino irradiation and 
nuclear reactions prior to shock arrival and (2) through the 
continuation of nuclear reactions induced by neutrinos as the 
stellar layers expand and cool. Neutrino nucleosynthesis is 
thought to be responsible for the production of, for example, 
$^{11}$B, $^{19}$F, and two of Nature's rarest isotopes: 
$^{138}$La and $^{180}$Ta. 

The production of the two isotopes, $^{11}$B and $^{10}$B, 
appears observationally to be linear with metalicity, i.e.,
primary mechanisms that operate early in the history of
our galaxy produce as much of these isotopes as secondary
(quadratic) mechanisms that operate after the Galaxy has
been enriched with metals. On the other hand, according to
current models, neutrino nucleosynthesis in
supernovae, which is a primary process, is not expected to
have produced much $^{10}$B, unlike the secondary process,
cosmic ray spallation. Thus, a laboratory calibration of 
the spallation channels producing these two isotopes that
can be used in conjunction with future HST observations 
discriminating between $^{10}$B and $^{11}$B would be 
invaluable in resolving this controversy and in supporting
the theory that neutrino nucleosynthesis in supernovae is
an important source of $^{11}$B in the Galaxy\cite{h99}.
$^{11}$B and $^{10}$B are produced through the following
spallation channels:

\begin{itemize}
\item $^{12}{\rm C}(\nu,\nu^{'}{\rm p})^{11}{\rm B}$
\item $^{12}{\rm C}(\nu,\nu^{'}{\rm n})^{11}{\rm C}(e^{+},\nu)^{11}{\rm B}$
\item $^{12}{\rm C}(\nu,\nu^{'}{\rm d})^{10}{\rm B}$
\item $^{12}{\rm C}(\nu,\nu^{'}{\rm pn})^{10}{\rm B}$
\end{itemize}

The final abundance of $^{19}$F produced in a supernova can serve as 
a ``supernova thermometer.'' If the abundance of $^{19}$F produced in 
the supernova is attributed to neutrino nucleosynthesis, the ratio of
[$^{19}$F/$^{20}$Ne]/[$^{19}$F/$^{20}$Ne]$_{\odot}$ (the denominator
is the measured ratio in the Sun) is a measure of the muon and tau 
neutrinosphere temperatures\cite{h99}.
$^{19}$F is produced through the following spallation channels\cite{h99}:

\begin{itemize}
\item $^{20}{\rm Ne}(\nu,\nu^{'}{\rm n})^{20}{\rm Ne^{*}}\longrightarrow
 ^{19}{\rm Ne}+{\rm n}\longrightarrow
 ^{19}{\rm F}+{\rm e}^{+}+\nu_{\rm e}+{\rm n}$
\item $^{20}{\rm Ne}(\nu,\nu^{'}{\rm n})^{20}{\rm Ne^{*}}\longrightarrow
 ^{19}{\rm F}+{\rm p}$
\end{itemize}

No obvious site for the production of the rare isotopes, $^{138}$La and 
$^{180}$Ta, has been proposed. That they can be produced via neutrino 
nucleosynthesis in supernovae is compelling, and may be very important in 
that their existence, however rare, may be a fingerprint of the neutrino 
process.
$^{138}$La, and $^{180}$Ta are produced through the following spallation 
channels:

\begin{itemize}
\item $^{139}{\rm La}(\nu,\nu^{'}{\rm n})^{138}{\rm La}$
\item $^{181}{\rm Ta}(\nu,\nu^{'}{\rm n})^{180}{\rm Ta}$
\end{itemize}

Experiments to measure the cross sections for all of these spallation channels
are being considered as part of a second wave of experiments at ORLaND.

\begin{center}
{\it The r-Process}
\end{center}

The site for the astrophysical r-process (rapid neutron
capture process) is not yet certain, but the leading candidate
is the neutrino-driven wind emanating from the proto-neutron
star after a core collapse supernova is initiated\cite{wmwhm94}. 
The r-process is thought to be responsible for roughly half of 
the Solar System's supply of heavy elements. As the neutrino-driven 
wind expands rapidly and cools, charged particle reactions ``freeze 
out'' while neutron capture reactions continue on the ``seed'' nuclei 
present at freeze-out. Neutron capture $(n,\gamma)$ reactions come 
into equilibrium with neutron disintegration $(\gamma,n)$ reactions 
as an equilibrium is established between the free neutrons and 
the nuclei in the wind. 
The $(n,\gamma)$--$(\gamma,n)$ equilibrium produces nuclei that
are quite neutron rich. Nuclei with half lives short compared to
the time scale for the r-process beta decay, producing nuclei
with higher Z and leading to the synthesis of heavier elements.
The simultaneous operation of these three types of reactions in
the wind and the accompanying nucleosynthesis constitutes the
r-process \cite{m94}.

Qian {\it et al.}~\cite{qhlv97} in both the $(n,\gamma)\leftrightarrow
(\gamma,n)$ equilibrium and the ``postprocessing phase'' after 
these reactions fall out of equilibrium have demonstrated that 
neutrino-induced reactions can significantly alter the r-process 
path and its yields. In the presence of a strong neutrino flux, 
$\nu_{\rm e}$-induced charged current reactions on the waiting 
point nuclei at the magic neutron numbers $N=50,82,126$ might 
compete with beta decays and speed up passage through the 
bottlenecks there. Also, neutrinos can inelastically scatter 
on r-process nuclei via $\nu_{\rm e}$-induced charged-current 
reactions and $\nu$-induced neutral-current reactions, 
leaving the nuclei in excited states that subsequently decay 
via the emission of one or more neutrons. This postprocessing 
may for example shift the abundance peak at $A=195$ to smaller 
mass. Taking things one step further, Haxton {\it et al.}~\cite{hlqv97}
pointed out that neutrino postprocessing effects would provide
a fingerprint of a supernova r-process. Eight abundances are
particularly sensitive to the neutrino postprocessing: 
$^{124}$Sn,
$^{125}$Te,
$^{126}$Te,
$^{183}$W,
$^{184}$W,
$^{185}$Re,
$^{186}$W, and
$^{187}$Re.
{\it Observed abundances of these elements are consistent with the
postprocessing of an r-process abundance pattern in a neutrino
fluence consistent with current supernova models.}

On a more pessimistic note, Meyer, McLaughlin, 
and Fuller\cite{mmf98} have investigated the impact of neutrino--nucleus 
interactions on the r-process yields and have discovered that electron neutrino 
capture on free neutrons and heavy nuclei (in the presence of a strong 
enough neutrino flux) can actually hinder the r-process by driving 
the neutrino-driven wind proton rich, posing a severe challenge to
theoretical models.

During the r-process and subsequent postprocessing in the 
supernova neutrino fluence, neutrinos interact with radioactive,
neutron-rich nuclei. Thus, relevant direct neutrino--nucleus 
measurements cannot be made. However, indirect measurements of 
charged- and neutral-current neutrino--nucleus interactions on 
stable nuclei that serve to gauge theoretical predictions would 
be invaluable.

\subsection{\bf Supernova Neutrino Detection}

The nineteen neutrino events detected by IMB and Kamiokande 
for SN1987A confirmed the basic supernova paradigm---that core
collapse supernovae are neutrino-driven events---and marked the
birth of extra-Solar-System neutrino astronomy. For a Galactic 
supernova, thousands of events will be seen by Super-K and SNO,
which, for the first time, will give us detailed neutrino 
``lightcurves'' and bring us volumes of information about 
the deepest regions in the explosion. In turn, these lightcurves
can be used to test and improve supernova models and their 
offshoot predictions. Moreover, comparing these detailed 
neutrino lightcurves with sophisticated supernova models 
could provide evidence for neutrino oscillations.

Among the neutrino--nucleus interactions of relevance for 
supernova neutrino detection are neutrino interactions on
deuterium in SNO, $^{16}$O in Super-K, and $^{56}$Fe and 
$^{206,207,208}$Pb in the proposed neutrino detector, OMNIS.

\begin{center}
{\it Deuterium: SNO}
\end{center}

The four main channels for supernova neutrino detection in
SNO are:
$\nu + {\rm e}^{-}\longrightarrow \nu + {\rm e}^{-}$,
$\nu + {\rm d}\longrightarrow \nu + {\rm p} + {\rm n}$,
$\nu_{\rm e} + {\rm d}\longrightarrow {\rm p} + {\rm p} + {\rm e}^{-}$, and
$\bar{\nu}_{\rm e} + {\rm d}\longrightarrow {\rm n} + {\rm n} + {\rm 
e}^{+}$.
Measurement of the reaction

\begin{itemize}
\item $\nu_{\rm e} + {\rm d}\longrightarrow {\rm p} + {\rm p} + {\rm e}^{-}$
\end{itemize}

\noindent at ORLaND, which is being considered to calibrate the reaction 
${\rm p} + {\rm p}\longrightarrow d + e^{+} + \nu_{\rm e}$ (part 
of the chain of reactions powering the Sun), would also provide a 
calibration of the SNO neutrino detector. Monte Carlo 
studies suggest that two years of data in approximately thirty 
fiducial tons of D$_{2}$O would yield a cross section measurement 
with an accuracy of a few percent\cite{workshop}, which, in turn, will enable a 
more accurate interpretation of the SNO data from 
the next Galactic supernova. The deuterium measurement is among 
the first wave of planned experiments at ORLaND.

\begin{center}
{\it Oxygen: Super-K}
\end{center}

The charged-current reaction $^{16}O(\nu_{\rm e},e^{-})^{16}F$ is 
the principle channel for electron neutrino interactions for thermal 
sources in the range $T_{\nu_{\rm e}}\geq 4-5$ MeV and its rate
exceeds that of neutrino--electron scattering by an order of magnitude 
for $T_{\nu_{\rm e}}\geq 7-9$ MeV\cite{h87}. Moreover, the electron 
angular distribution is strongly correlated with the electron neutrino 
energy, providing a way to measure the incident neutrino energy and, 
consequently, the electron neutrino spectra. By inference, one would 
then be able to measure, for example, the electron neutrinosphere 
temperature in a core collapse supernova, i.e., we would have a 
supernova thermometer\cite{workshop}. In addition, the appearance of back-angle 
electron emission from this reaction in, for example, Super-K would 
result from very energetic electron neutrinos, more energetic 
than predicted by supernova models. This would be evidence for flavor 
oscillations\cite{workshop}. Muon and tau neutrinos in the stellar core couple to the 
core material only via neutral currents, whereas electron neutrinos and 
antineutrinos couple via both neutral and charged currents. As a result, 
the former decouple at higher density and, therefore, temperature, and 
have harder spectra. {\it In fact, terrestrial detection of the tau and 
electron neutrinos from a Galactic or near--extra-Galactic core collapse 
supernova may be our only hope of ever observing oscillations between 
these two neutrino flavors if the mixing angle is small.} 

Utilizing reactions on $^{16}$O, Langanke, Vogel, and Kolbe\cite{lvk96} have 
suggested a novel way of also unambiguously identifying muon and tau neutrino 
signatures in Super-K. The large average energies for these neutrino flavors 
are sufficient to excite giant resonances via the neutral-current reactions 
$^{16}O(\nu_{\mu,\tau},\nu_{\mu,\tau}^{'})^{16}O^{*}$.
These resonances are above particle threshold and subsequently decay via the 
emission of protons, neutrons, and gamma rays. The gamma rays would provide 
the muon and tau neutrino signatures. The two decay channels are:
$^{16}O^{*}\longrightarrow ^{15}O + {\rm n} + \gamma$ and
$^{16}O^{*}\longrightarrow ^{15}N + {\rm p} + \gamma$.

Thus, accurate measurements of both charged- and neutral-current 
neutrino cross sections on $^{16}$O would be foundational to 
interpreting the neutrino data from the next Galactic core collapse supernova 
and to using that data to potentially observe, for the first time,
flavor oscillations involving the tau and electron neutrinos.

An experiment to measure the cross section for:

\begin{itemize}
\item $^{16}O(\nu_{\rm e},e^{-})^{16}F$
\end{itemize}

\noindent is among the first proposed experiments at ORLaND. 
Future experiments may focus on the cross sections 
for:

\begin{itemize}
\item $^{16}O(\nu_{\mu},\nu_{\mu}^{'}{\rm n}\gamma)^{15}O$
\item $^{16}O(\nu_{\mu},\nu_{\mu}^{'}{\rm p}\gamma)^{15}N$
\end{itemize}

\begin{center}
{\it Iron and Lead: OMNIS}
\end{center}

The use of iron and lead in OMNIS would provide yet another way 
of measuring neutrino oscillations in core collapse supernovae\cite{b00}.
Iron has a sufficiently high threshold for neutron production
via charged-current neutrino interactions that such production is 
negligible, whereas, in lead, neutrons are produced by both charged- 
and neutral-current interactions. Oscillations between the more 
energetic muon and tau neutrinos and the electron neutrinos would boost
the charged-current event rate while leaving the neutral-current 
rate roughly unchanged. Thus, the ratio of the event rate in lead 
to that in iron would serve as an indicator that oscillations 
had occurred.

To develop OMNIS, experiments to measure the neutrino--iron and 
neutrino--lead cross sections at ORLaND have been proposed. For 
iron, the neutral-current reaction:

\begin{itemize}
\item $^{56}{\rm Fe}(\nu,\nu')^{56}{\rm Fe}^{*} \longrightarrow ^{55}{\rm Fe}+{\rm n}$
\end{itemize}

\noindent dominates. For lead, a total cross section would be measured 
resulting from the following neutral- and charged-current channels:

\begin{itemize}
\item $^{208}{\rm Pb}(\nu,\nu')^{208}{\rm Pb}^{*} \longrightarrow ^{207}{\rm Pb}+{\rm n}$
\item $^{208}{\rm Pb}(\nu,\nu')^{208}{\rm Pb}^{*} \longrightarrow ^{206}{\rm Pb}+{\rm n}+{\rm n}$
\item $^{208}{\rm Pb}(\nu_{\rm e},e^{-})^{208}{\rm Bi}^{*} \longrightarrow ^{207}{\rm Bi}+{\rm n}$
\end{itemize}

\noindent including the channels for the isotopes $^{206}{\rm Pb}$ 
and $^{207}{\rm Pb}$. The iron and lead cross section measurements
are among the first proposed experiments at ORLaND.

\begin{figure}
\begin{center}
\epsfysize=2.5in 
\epsfbox{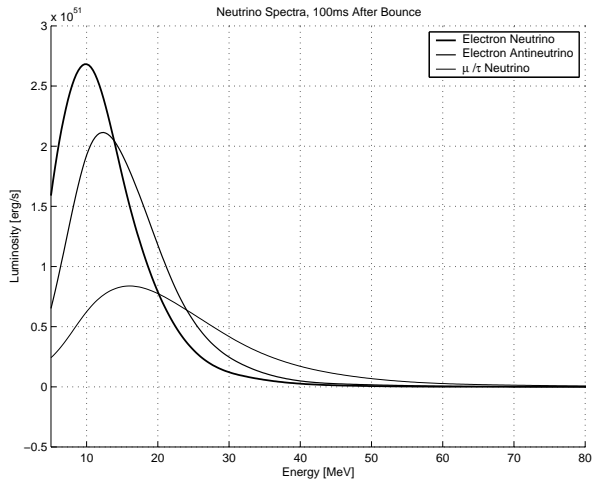} 
\caption[h]{Neutrino spectra from a supernova simulation with Boltzmann 
neutrino transport initiated from a 13 M$_{\odot}$ progenitor. The 
simulation is fully general relativistic, and the spectra are computed 
at a radius of 500 km\cite{lmtmhb00}.
}
\end{center}
\end{figure}

\begin{figure}
\begin{center}
\epsfysize=2.5in 
\epsfbox{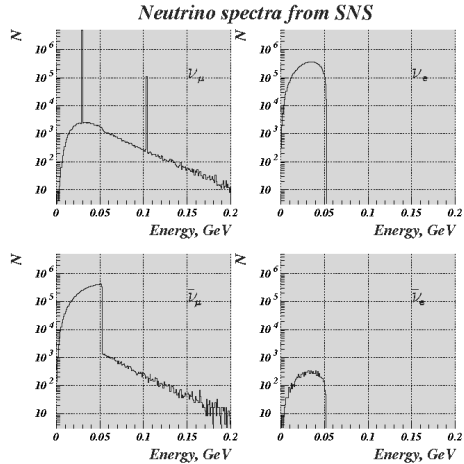} 
\caption[h]{SNS neutrino spectra.}
\end{center}
\end{figure}

\section{ORLaND}
A new facility to measure neutrino--nucleus cross sections, such as the proposed ORLaND facility 
at Oak Ridge, would provide an experimental foundation for the many 
neutrino--nucleus weak interaction rates needed in supernova models. Indeed, 
we are presented with a unique opportunity, given the intensity of the SNS 
as a neutrino source and given the overlap (shown in Figures 5 and 6) between 
the spectra of SNS and supernova neutrinos, to make such measurements. This 
would enable more realistic supernova models and allow us to  
cull fundamental physics from these models with greater confidence 
by comparing them with 
detailed observations. 
Charged- and neutral-current neutrino interactions on nuclei in the stellar 
core play a central role in supernova dynamics, nucleosynthesis, and neutrino 
detection. Measurements of these reactions on select, judiciously chosen targets 
would provide an invaluable test of the complex theoretical models used to compute 
the neutrino--nucleus cross sections.

\section*{Acknowledgments}
A.M. is supported at the Oak Ridge National Laboratory, managed by
UT-Battelle, LLC, for the U.S. Department of Energy under contract
DE-AC05-00OR22725. AM would like to acknowledge many illuminating
discussions with Frank Avignone, John Beacom, Jeff Blackmon, Dick 
Boyd, David Dean, Yuri Efremenko, Jon Engel, George Fuller, Wick 
Haxton, Raph Hix, Ken Lande, Karlheinz Langanke, Matthias Liebend\"{o}rfer,
Gabriel Martinez-Pinedo, Gail McLaughlin, Mike Strayer, and Friedel 
Thielemann, all of which contributed significantly to this manuscript.

\end{document}